\documentclass[twocolumn,a4paper]{IEEEtran}
\usepackage{graphics}
\usepackage{graphicx}
\usepackage{amsmath}
\newcommand{\x}{\xi}
\newcommand{\D}{{\cal D}}
\newcommand{\w}{\omega}
\newcommand{\mvoice}{${\cal V}$}
\newcommand{\mmusic}{${\cal M}$}
\newcommand{\voice}{{\cal V}}
\newcommand{\music}{{\cal M}}
\newcommand{\freqone}{240}
\newcommand{\freqtwo}{738}
\newcommand{\freqthree}{1361}
\newcommand{\bwone}{200}
\newcommand{\bwtwo}{606}
\newcommand{\bwthree}{1246}
\newcommand{\fone}{1 (center frequency $\w_0 = \freqone$)}
\newcommand{\ftwo}{2 (center frequency $\w_0 = \freqtwo$)}
\newcommand{\fthree}{3 (center frequency $\w_0 = \freqthree$)}
\newcommand{\vsegments}{$150$}
\newcommand{\msegments}{$100$}
\newcommand{\tsegments}{$250$}
\newcommand{\vrsegments}{$30$}
\newcommand{\mrsegments}{$20$}
\newcommand{\vtsegments}{$120$}
\newcommand{\mtsegments}{$80$}

\begin{document}

\title{Music and Vocal Separation Using Multi-Band Modulation Based Features}

\author{
\IEEEauthorblockN{Sunil Kumar Kopparapu, Meghna Pandharipande, G Sita} \\
\IEEEauthorblockA{
TCS Innovation lab - Mumbai, 
Tata Consultancy Services Limited,
Yantra Park, Thane (West) - 400 601.    \\
Email: \{SunilKumar.Kopparapu, Meghna.Pandharipande\}@TCS.Com}
%Sita.UB@Gmail.Com }
}

\maketitle
\thispagestyle{empty}

\begin{abstract}
The potential use of non-linear speech features has not 
been investigated for music analysis although other commonly used speech features like
Mel Frequency Ceptral Coefficients (MFCC) and pitch have been used extensively. 
In this paper, we assume an audio signal to be a sum of 
modulated sinusoidal 
and then use the energy separation algorithm to decompose the audio into 
amplitude and frequency modulation components 
using the non-linear Teager-Kaiser energy operator. 
We first identify the distribution of these non-linear 
features for music only and voice only segments in the audio signal in
different Mel spaced frequency bands and show that they have the ability to
discriminate.
The proposed method based on Kullback-Leibler divergence 
measure is evaluated using a set of Indian classical songs from three 
different artists. Experimental results show that the discrimination 
ability is evident in certain low and mid frequency bands ($200$ - $1500$ Hz). 

\end{abstract}


\section{Introduction}
 
Increased availability and use of large digital music corpora, requires intelligent 
music management systems. This has  resulted in the development of a variety of 
intelligent music content management systems.  Automatic segmentation of song into 
vocal and music regions is a very important step for several applications 
like singer 
identification, musical instrument analysis, preference based searches of music, 
to name a few. Audio search, annotation and browsing applications also benefit 
greatly with automatic music voice segmentation. Early work in this area 
includes the work carried out by Berenzweig and Ellis \cite{ellis} where 
they suggest the use of an artificial neural network (ANN) to train on radio 
recordings to segment songs into vocal and non-vocal (music) regions. 
Kim and Whitman \cite{kim} filter the audio signal using a bandpass 
filter and then used hormonicity as a measure to detect vocal 
regions and separate them from music. 
%--Meghna--.
More recently, Sridhar and Geetha \cite{Sridhar}, identified swaras in South Indian
classical music by finding the pitch for a particular segment which in turn gives 
information about middle octave swara. Ramona et al \cite{Mathieu} 
use support vector machine to separate singing voice from pure
instrumental region while  Demir and others \cite{Demir} 
use hidden Markov model (HMM) 
based acoustic models to calculate posterior probabilities to segment an 
audio signal as speech and music. Zhou \cite{Zhou}
discriminates voice and music using novel spectral feature like averaged cepstrum. 
and Kos et al \cite{Kos} on-line segment speech music 
for broadcast news domain using Mel-Frequency Cepstral Coefficients Variance 
(MFCCV).
Barbedo and others \cite{Barbedo} propose a mechanism to 
discriminate speech and music signal by extracting four features and 
then combining them linearly into a unique parameter. Didiot \cite{Didiot}
propose a wavelet based signal decomposition instead of Fourier Transform for 
discriminating speech and music.
In most of the work cited above the use of conventional speech features is
apparent because of processing audio using linear system theory. In 
this paper, we   propose and investigate the use of non-linear feature set to
discriminate speech and music.  Non-linear speech models 
attempt to model the spectral variability of the speech signal and decompose it 
into 
amplitude modulation (AM) and frequency modulation (FM) components.
%--Meghna--.
Modulation of the amplitude and/or frequency of a sine wave has been used 
extensively 
in communication systems for transmitting information \cite{quatieri},
\cite{book_01}. 
%The recent state 
%of basic knowledge on amplitude-modulation (AM) and frequency-modulation (FM) systems 
%can be found in contemporary books on communications,e.g \cite{book_01} 

Use of nonlinear analysis for speech processing \cite{DBLP:conf/nolisp/2005}
has of late received attention specifically for speaker recognition, speech
analysis, voice pathologies, speech recognition and speech enhancement.
Specifically, decomposing a non-stationary, bandpass signal into its AM and FM 
components has been 
addressed by many researchers and a number of techniques have been published in 
literature. The most popular approaches are based on the auditory motivated 
decomposition proposed by Quatieri et al \cite{quatieri} and Teager energy 
based algorithms proposed by Dimitrios et al \cite{maragos1}. Features derived 
using non-linear speech framework could reveal the potential of alternative 
speech models in various speech applications such as speaker identification 
\cite{wenn}, vocal fold pathology assessment \cite{john} and even emotion
classification \cite{10.1109/SNPD.2007.487}. Dimitrios et 
al \cite{maragos} have used these AM-FM features for phoneme classification 
and speech recognition tasks \cite{dimi_05}. They concluded that these 
non-linear speech features could be efficiently used in speech 
classification and recognition tasks.

In this paper, we use Teager energy based algorithm \cite{maragos1} to obtain 
modulation based features from an audio stream. These features are then 
used in a 
supervised learning scheme for segment-wise discrimination of vocal and music
component in an audio stream. We first analyze the signal to obtain the 
instantaneous frequency distribution in a number of Mel scale signified 
frequency bands for the entire duration of the audio stream.  The task of 
identifying an audio segment as being vocal or music is determined by 
measuring the well known Kullback-Leibler divergence between the distributions 
of reference music and vocal regions with the corresponding distributions of 
the test audio stream. The rest of the paper is organized as follows: Section
\ref{sec:modulation_features} explains the theory of  non-linear speech 
modeling and feature extraction method, Section \ref{sec:experiments} 
explains about the data and the methodology used in our experiments. 
We conclude in Section \ref{sec:conclude}.

\section{Modulation based feature extraction}
\label{sec:modulation_features}
%--Meghna--.
Audio signal $x(t)$ is non-linear, time-varying and can be looked upon as a
AM-FM model as follows (as mentioned in \cite{maragos1})
\begin{equation}
%x(t) = a(t) \cos \left ( \int_{0}^{t}\w(\tau)d\tau \right )
x(t) = a(t) \cos \left ( \phi(t)\right )
\label{eq:am_fm_rep_0}
\end{equation}
where,  $a(t)$ is the time varying amplitude and $\phi(t)$ is defined as
\begin{equation}
\phi\left (t  \right )= \w_{c} t + \w _{m}\int_{0}^{t}q\left 
(\tau   \right )d\tau +\theta
\label{eq:phit}
\end{equation}
where $\w_{c}$ is the center frequency and $\w _{m}$ is the maximum
frequency deviation from the $\w_{c}$,
 $ \left|q\left(t\right)\right|\leq 1 $  
and $\theta = \phi \left (0 \right)$ is some arbitrary constant phase offset. 
The time varying instantaneous angular frequency $\w _{i}$ id defined as
\begin{equation}
\w _{i}\left ( t \right )\stackrel{def}{=} \frac{d}{dt} \phi\left (t  \right )=
\w_{c} + \w_{m} q\left (  t\right )
\label{eq:ifreq} \end{equation}

Note that (\ref{eq:am_fm_rep_0}) has both an AM and FM structure
hence we call $x(t)$ an AM-FM signal. 
It has 
been shown that this non-linear modeling of speech helps in extraction 
of robust features for speech \cite{dimi_05}.
These features measure the amount of amplitude and 
frequency modulation that exists in the audio signal and attempt to model 
aspects of speech acoustic information.
Further, two different information 
signals can be simultaneously transmitted in the 
amplitude  $a(t)$ 
and the frequency $ \w_{i}(t)$. Such AM-FM signals 
are very frequently used in communication systems \cite{quatieri}.
The AM-FM model can be used to represent a speech signal 
$s(t)$ as a sum of AM-FM signals, namely
\begin{equation}
s(t) = \sum_{k=1}^{K} a_{k}(t)cos (\phi _{k}(t))
\label{eq:sig_in_am_fm_rep}
\end{equation}
where $K$ is the number of speech formats. 
Clearly, $a(t)$ and $\w_i(t)$ for $k = 1, 2, \cdots K$ represents the
speech signal $s(t)$. Now, 
given a speech signal over some time 
interval, the problem is to estimate the amplitude envelope $\left |a(t)\right |$ 
and the instantaneous frequency $\w_{i}\left ( t \right )$ 
of each $k$ and at each time $t$. 
One of the ways to estimate $a(t)$ and $\w_i(t)$, is to first 
isolate individual resonance by bandpass filtering the speech signal 
around its formants and then  estimating 
amplitude and frequency modulating signals of each resonance based 
on an "energy-tracking operator" as described in \cite{quatieri}.
The Teager energy operator $\psi$ (TEO) is defined as,
\begin{equation}
\psi_{c}[x(t)] \stackrel{def}{=} \left [\frac{d}{dt} x(t) \right ]^{2} 
- x(t) \left [ \frac{d^2}{dt^2} x(t) \right ]
\label{eq:teager_operator_1}
\end{equation}
When $\psi$ given by (\ref{eq:teager_operator_1}) is applied to the 
bandpass filtered speech signal 
(\ref{eq:am_fm_rep_0}), we get the instantaneous source energy, namely, 
\begin{equation} 
\psi[x(t)] \approx a^2(t)\w_{i}^{2}\left( t \right )
\label{eq:instant_energy} 
\end{equation} 
%The continuous form of the Teager Energy Operator TEO is shown in 
%(\ref{eq:teager_operator}). 
In the discrete form as is applicable to most speech processing systems 
\cite{Guojun} (\ref{eq:teager_operator_1}) can be written as 
%the operator for discrete-time signal from its continious form 
%$\psi_{c} =\left [ x\left ( t \right ) \right ]$, as
\begin{equation} 
\psi \left [ x\left [ n \right ] \right ]= x^{2}\left 
[n\right ]-x\left [ n+1 \right ]x\left [n-1\right ]
\label{eq:teager_operator_2}
\end{equation}
where $x[n]$ is the sampled speech signal.

%The TEO is typically applied to a bandpass filtered speech signal, since 
%its intent is to reflect the energy of the nonlinear flow within the vocal 
%tract for a single resonant frequency. Although the output of the bandpass 
%filter still contains more than one frequency component, it can be 
%considered as an AM-FM signal,
%$r\left ( t \right )=a\left ( t \right )cos\left (2\pi f\left (t\right )
%t\right )$. The TEO output of $r\left ( t \right )$ can be approximated as
%\begin{equation} 
%\psi \left [ r\left (t  \right ) \right ]\approx \left [ a\left 
%( t \right ) 2\pi f\left ( t \right )\right ]^{2}
%\label{eq:teager_operator}
%\end{equation}
The AM-FM demodulation can be achieved by separating the instantaneous 
energy given in (\ref{eq:instant_energy}) into its amplitude and frequency 
components. $\psi$ is the main ingredient of the first Energy Separation 
Algorithm (ESA) developed in \cite{maragos1} and used for signal 
and speech AM-FM demodulation.
\begin{equation}
f[n] \approx %\frac{1}{2\pi T}
\cos^{-1}\left (1 - \frac{\psi\left [y[n]  \right ]+\psi \left [ y[n+1] \right ] }
{4\psi \left [x[n]  \right ]} \right )
\label{eq:fm}
\end{equation} and
\begin{equation}
\left | a[n] \right |\approx \sqrt{\frac{\psi \left [x[n] \right ]}
{\left [1-\left ( 1 - \frac{\psi\left [y[n]  
\right ]+\psi \left [ y[n+1] \right ] }{4\psi \left [x[n]  \right ]} \right )^{2}  \right ]}}
\label{eq:am}
\end{equation}
\newline
where
$y[n] = x[n] - x[n-1]$  and 
$f[n]$ is the FM component at sample $n$
and
$a[n]$ is the AM component at sample $n$. 
In practice the speech signal is bandpass filtered using Gabor filters because
of their optimal time-frequency discriminability \cite{maragos1}, namely,
%The ESA cannot handle wideband signals, such as speech signals, due 
%to inherent limitations of the algorithm. One efficient way to deal with 
%uch limitations is bandpass filtering of the signal. A single AM-FM signal 
%can be assumed to be present if the audio signal is bandpass filtered using 
%a sufficiently narrow-band filter. For this purpose, the Gabor filters are 
%chosen for several reasons listed in \cite{maragos1}, such as 
%optimal time-frequency discriminability. The continious TEO 
%(Teager Energy Operator) $\psi$, is combined with bandpass filtering 
%and sampled at time instances $t = nT$ is given by,
\begin{equation} 
s\left (t\right )= x\left (t\right )*g\left (t\right)
\label{eq:gabor_1}
\end{equation} 
where $g(t)$ is given by,
\begin{equation}
g(t) = \frac{1}{\sqrt{2 \pi} \sigma} \exp^{-\frac{t^2}{2 \sigma^2}}
\exp^{i(2\pi\w_0 t)}
\label{eq:gabor_2}
\end{equation}
where $\w_0$ is the center frequency and $\sigma$ is the 
bandwidth of the Gabor filter. In the case of speech or audio signals, 
a Gabor filter-bank (placed at various critical band frequencies such 
as formant frequencies or at frequencies determined by Mel-scale) 
with a narrow bandwidth are used. The extraction of AM-FM components 
(\ref{eq:fm}) and (\ref{eq:am}) from the bandpass filtered signal 
may be carried out 
using the Teager energy of the filtered signal.
The efficiency of non-linear speech features, namely instantaneous 
modulation features such as instantaneous amplitude and instantaneous 
frequencies around different resonance frequencies of the speech signal 
have been studied for various applications in speech processing area such 
as phoneme classification, speech recognition \cite{maragos, dimi_05}, 
assessment of vocal fold pathology \cite{john}, stress detection \cite{mandar}. In this paper, we investigate the performance of instantaneous frequency 
modulation features to automatically discriminate vocal and music dominant 
regions in an audio track. The hypothesis that the instantaneous 
modulation feature distribution may be different for vocal and music dominant 
regions is derived from the observation that the generative sources of voice 
and music are different. 
For instance, the singing voice (vocal) is replete with 
large pitch modulations unlike the music component.
Additionally, the voice harmonics 
in the spectrum are observed to be below $5$ kHz where as the music energy 
is observed to be spread throughout the 
spectrum up to $10$ kHz with certain frequency bands dominated entirely 
by the music energy.

\section{Experimental results and discussion}
\label{sec:experiments}

\begin{figure}
\includegraphics[width=0.45\textwidth] {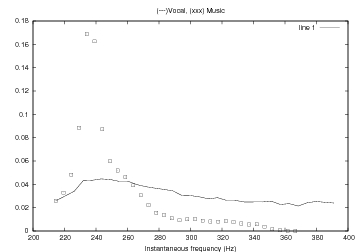}
\caption{Comparison of distribution of Instantaneous frequencies 
in band \fone\ for \mvoice\ and \mmusic.}
\label{multi1}
\end{figure}

\begin{figure}
\includegraphics[width=0.45\textwidth]{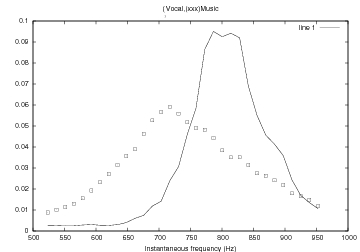}
\caption{Comparison of distribution of
Instantaneous frequencies in band \ftwo\ for \mvoice\ and \mmusic.}
\label{multi2}
\end{figure}

\begin{figure}
\includegraphics[width=0.45\textwidth]{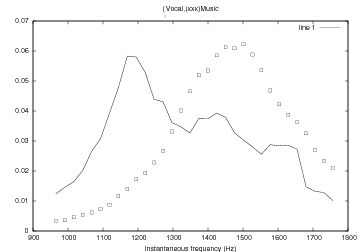}
\caption{Comparison of distribution of
Instantaneous frequencies in band \fthree\ for \mvoice\ and \mmusic.}
\label{multi3}
\end{figure}

\begin{figure}
\includegraphics[width=0.45\textwidth]{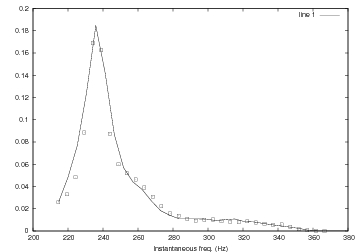}
\caption{Instantaneous
frequency distribution in band \fone\ for two typical \mvoice\ segments}
\label{multi4}
\end{figure}

\begin{figure}
\includegraphics[width=0.45\textwidth]{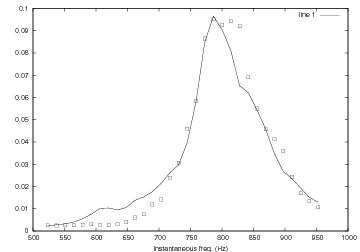}
\caption{Instantaneous
frequencies distribution in  band \ftwo\ for two typical \mmusic\ segments.}
\label{multi5}
\end{figure}

\begin{figure}
\includegraphics[width=0.45\textwidth]{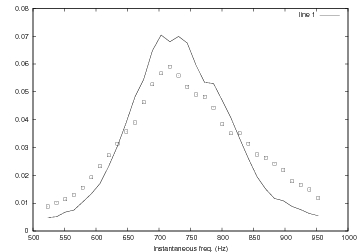}
\caption{Instantaneous frequency distribution in band \ftwo\ for two 
typical \mvoice\ segments.}
\label{multi6}
\end{figure}

To identify and test the performance of 
non-linear features for voice music separation, we collected several 
audio signals which had portions of speech (or voice) and music. Distinct 
voice
and music   are an essential part of Indian classical music; so we collected a
large set of Indian classical music and stored them in {\em wav} format.
The collection database consisted of a total length of approximately $475$ 
s of audio stream. The audio was sampled at $22.05$ kHz and was manually 
labeled as \mvoice\ (for voice) or \mmusic\ (for music) using a semi-automatic
process and later manually checked for the correctness of classification.
We use the Mel spaced Gabor filter-bank \cite{maragos}
to filter the audio into the first three bands. For each of these 
filtered signal we computed the  non-linear instantaneous features. 
We restricted our analysis to the lower three 
filter bands as in our preliminary investigations, we found that the 
discrimination power to segment voice and music is evident in these 
three bands. %\footnote{We are in the process of experimenting if other 
%bands canimprove the performance}. 
Further it was observed that the instantaneous amplitudes in 
various bands are not discriminative enough 
for different audio segments. Hence in all our experiments we have not
considered instantaneous amplitudes. 
The three Mel-spaced center 
frequencies ($\w_0$ in (\ref{eq:gabor_2})) we have analyzed are, 
$\freqone$ Hz, $\freqtwo$ Hz and $\freqthree$ Hz with a bandwidth 
($\sigma$ in (\ref{eq:gabor_2})) of $\bwone$ Hz, $\bwtwo$ Hz and $\bwthree$ Hz 
respectively. 
These reference audio segments which have been tagged as \mmusic\ and \mvoice\ are first bandpass filtered 
using Gabor filter-bank (\ref{eq:gabor_2}) at three different $\w_0$'s namely,
$\freqone$, $\freqtwo$ and  $\freqthree$. 
%mentioned in Section \ref{sec:modulation_features}. 
Instantaneous frequency components are obtained for each of the 
filtered signal using (\ref{eq:fm}) and (\ref{eq:am}).
% the Teager energy based discrete energy 
%separation algorithm. 
Figure \ref{multi1} to Figure \ref{multi6} show the typical 
distributions for vocal (\mvoice) and music (\mmusic) segments of the audio. We 
took the audio signal and extracted the instantaneous frequencies 
using three different Gabor filters. We then segregated the 
instantaneous frequencies based on the tags, namely, \mvoice\ and \mmusic. 
These tags are used to get the instantaneous feature distribution for 
voice and music segments of the audio, for three different bands, 
namely, band \fone, band \ftwo\ and band \fthree\ are  shown 
in Figure \ref{multi1}, 
Figure \ref{multi2} and Figure \ref{multi3} respectively.


It can also be seen from the distribution plots that the instantaneous 
frequency has a very  distinct distributions for voice and music segments 
in all the three frequency bands. Additionally, the instantaneous 
frequency distribution of similar tags (namely, for \mvoice\ and \mmusic) 
show similar distribution (see Figures \ref{multi4}, \ref{multi5} and 
\ref{multi6}). 
This observation suggested that non-linear speech parameters {\em do} 
have the ability to distinguish voice and music components very reliably.

In all our experiments we use the Kullback-Leibler (KL) divergence metric,
namely,
\begin{equation}
\D(p_1(\x), p_2(\x))= \int p_1(\x) \log \left ( \frac{p_1(\x)}{p_2(\x)} \right
) d\x
\label{eq:kl}
\end{equation}
to compare the distance between any two distributions. 
If $p_1(\x)$ and $p_2(\x)$ 
are
two distributions then the distance of the distribution $p_2(\x)$ 
from $p_1(\x)$  is given by (\ref{eq:kl}).

In all we had \tsegments\ voice and musics segments 
of which \msegments\ segments had the tag \mmusic\ and
\vsegments\ segments had the tag \mvoice. Is segment was of an average duration
of $2$ s. % 475/250
The reference instantaneous frequency distribution for the  voice and music 
segments of the 
audio signal is computed from  the reference vocal and music segments for all
the three bands. 
The reference is created using $20$\% of the segments in each category, 
namely \mrsegments\ segments for
\mmusic\ and \vrsegments\ segments for \mvoice, and the rest namely
\mtsegments\ segments of \mmusic\ and \vtsegments\ segments of \mvoice\ which
were not part of the reference segments were used to test the performance of
the proposed approach to recognize \mmusic\ and \mvoice.  
Using 
these audio reference ( \mrsegments\ segments for
\mmusic\ and \vrsegments\ segments for \mvoice) 
we construct the distributions $p_{\music}$ 
and $p_{\voice}$. A test audio segment ($T$) is taken (which is not 
part of the audio that has been used to create the reference) 
and the instantaneous frequency distribution of $T$ is computed, 
as $p_T$. The distribution of the test segment ($p_T$) is compared 
with the distribution of the reference music $p_{\music}$ and reference 
vocal $p_{\voice}$ distributions using (\ref{eq:kl}). 
Namely, we compute $\D(p_{\voice}, p_T)$ and $\D(p_{\music},  p_T)$. 
If $\D(p_{\voice}, p_T) < \D(p_{\music},  p_T)$ the $T$ is classified 
as $\voice$, else $T$ is classified as $\music$. 
%The closeness of the test audio segment
%distribution with the reference segment determines the 
%audio segment.
%Hence a comparison of these distributions is carried out for various 
%segments in order to classify the segment as vocal or music segment 
%using Kullback-Leibler divergence which is given as,

A $5$ fold cross
validation was used to arrive at the performance of using non-linear features
to discriminate voice and speech.
Table \ref{tab:results} tabulates the $5$ fold cross
validation experimental results.
\begin{table}
\begin{center}
\begin{tabular}{|c|c||c|c|c|} 
\hline
 & {Number of }   & {Number of }  & {Number of } \\  
 & {Segments}   & {Misrecognised} & {Correct Recognitions}   \\ \hline 
Vocal (\mvoice)  & \vsegments\  &  22  & 128 (85 \%)\\ \hline 
Music (\mmusic)  & \msegments\   & 12  &  78 (78\%)\\ \hline 
\end {tabular}
\end {center}
\caption{Voice and Music segmentation results with $5$ fold cross validation.}
\label{tab:results}
\end {table}
As it can be seen the use of non-linear features for segmentation 
of voice and music is able to segment music and voice quite well. 
The use of MFCC as the features resulted in large number of 
misrecognitions as compared to the misrecognitions dues to 
non-linear instantaneous frequency features.

\section{Conclusions}
\label{sec:conclude}

Use of non-linear speech features has not been used for music and voice
classification though it has been used in some areas of speech recognition,
speaker identification. In this paper we have used the instantaneous frequency
calculated over band filtered speech signal to discriminate speech and voice.
We first assumed a sum of modulated sinusoidal model for audio signal 
and investigated the performance of instantaneous frequency 
modulation feature in discriminating voice and music segments.
We used Gabor filters to restrict  the analysis to a limited 
number of carrier frequencies which are nothing but the center frequencies 
of the bandpass filters. We first observed that the distribution of the 
instantaneous frequency
feature  over three bands (centered at \freqone, \freqtwo\ and  \freqthree) 
is able to discriminate voice and music.
This observation was   exploited 
to classify the audio stream into music and voice segments. 
Future work would involve 
extensive testing of the method with larger data-set for consistency of the 
results and testing across various genres of music.

\section*{Acknowledgment}
The authors would like to thank the members of the TCS Innovation Labs - Mumbai
for the great working environment.

%\newpage
\bibliographystyle{IEEEtran}
\bibliography{voice_music}

\end{document}